\documentclass[manuscript, noblind]{geophysics}

\usepackage{multirow}
\usepackage{amsmath}

\usepackage[colorlinks,
linkcolor=blue,
urlcolor = blue,
citecolor=blue
]{hyperref}

\begin{document}

\title{High-resolution closed-loop seismic inversion network in time-frequency phase mixed domain}

\renewcommand{\thefootnote}{\fnsymbol{footnote}} 


\address{
	\footnotemark[1]Key Laboratory of Earth Exploration and Information Technology of Ministry of Education, Chengdu University of Technology, China. E-mail: yingtianliu06@outlook.com\\
	\footnotemark[2]State Key Lab of Oil and Gas Reservoir Geology and Exploitation, Chengdu University of Technology, China. E-mail: liyong07@cdut.edu.cn (corresponding author); 2390328914@qq.com; zhangquanliao@outlook.com; 454810391@qq.com;\\
}

\author{Yingtian Liu\footnotemark[1], Yong Li\footnotemark[2], Junheng Peng\footnotemark[2], Huating Li\footnotemark[2], Mingwei Wang\footnotemark[2]}

\lefthead{Liu et al.} 
\righthead{The TFP-CSIN} 

\maketitle

\begin{abstract}
Thin layers and reservoirs may be concealed in areas of low seismic reflection amplitude, making them difficult to recognize. Deep learning (DL) techniques provide new opportunities for accurate impedance prediction by establishing a nonlinear mapping between seismic data and impedance. However, existing methods primarily use time domain seismic data, which limits the capture of frequency bands, thus leading to insufficient resolution of the inversion results. To address these problems, we introduce a new time-frequency-phase (TFP) mixed-domain closed-loop seismic inversion network (TFP-CSIN) to improve the identification of thin layers and reservoirs. First, the inversion network and closed-loop network are constructed by using bidirectional gated recurrent units (Bi-GRU) and convolutional neural network (CNN) architectures, enabling bidirectional mapping between seismic data and impedance data. Next, to comprehensive learning across the entire frequency spectrum, the Fourier transform is used to capture frequency information and establish frequency domain constraints. At the same time, the phase domain constraint is introduced through Hilbert transformation, which improves the method's ability to recognize the weak reflection region features. Both experiments on the synthetic data show that TFP-CSIN outperforms the traditional supervised learning method and time domain semi-supervised learning methods in seismic inversion. The field data further verify that the proposed method improves the identification ability of weak reflection areas and thin layers. 
\end{abstract}

\section{Introduction}

Seismic inversion plays a critical role in stratigraphic interpretation because it can estimate the rock properties of the entire seismic profile \citep{mustafa2021joint}. In this technique, the spatial structure and physical properties of subsurface rock formations are imaged and solved by using seismic data combined with known geological laws and drilling and logging data as constraints \citep{pei2021high}. As an important seismic inversion method, seismic impedance inversion directly reflects the nature of subsurface rocks and can effectively indicate rock characteristics and reservoir information \citep{lin2023time}. However, due to the incompleteness of the observed data, noise, and the limitation of the wavelet band, the inversion equations are often ill-posed \citep{wang2010seismic,tamaddon2020data,liu2024anti}. To solve these problems, researchers have proposed a variety of nonlinear inversion methods, including the simulated annealing method \citep{mosegaard1991simulated,srivastava2009fractal}, ant colony optimization algorithm \citep{conti2013fast}, and genetic algorithm \citep{nolte1994vertical,maurya2023qualitative}. These methods provide new ideas and solutions to improve the accuracy and reliability of seismic impedance inversion. Although these seismic inversion techniques have been continuously improved, they still face certain limitations: reliance on a limited number of geophysical parameters, constraints imposed by mathematical physical models, high computational demands, and partial dependence on subjective expertise \citep{farfour2015seismic,kemper2014joint,madiba2003seismic,riedel2009acoustic}. Therefore, the effect of the impedance inversion method is not satisfactory in practical applications, especially in reservoir prediction and lithology characterization \citep{she2019seismic,zhou2021robust,zhou2022bayesian}.

With the increase of computing power, deep learning (DL) methods have attracted wide attention and have been successfully implemented in many fields, including computer vision \citep{ioannidou2017deep,brunetti2018computer}, natural language processing \citep{chowdhary2020natural,zhang2024survey} and image recognition \citep{wang2018successful,li2023modified}. DL methods have also been successfully applied in the field of geophysical exploration. Researchers have combined deep neural networks with specific problems in seismic exploration, including first break picking \citep{yuan2018seismic,yin2023first}, seismic fault identification \citep{lin2022automatic,li2024fault}, seismic horizon tracking \citep{he2023multiple,luo2023sequence}, and seismic phase classification \citep{saad2021capsphase,feng2022edgephase}.

In recent years, with the optimization and development of DL network structures, seismic impedance inversion has undergone significant technical innovations. Various network architectures have been employed to address the challenges in seismic impedance inversion. \citet{das2019convolutional} used one-dimensional convolutional neural network (CNN) to train inverse mapping models, demonstrating the considerable potential of CNN in impedance prediction. \citet{mustafa2019estimation} employed temporal convolutional networks to predict impedance from seismic data, addressing the issues of vanishing gradients in recurrent neural networks and overfitting in CNN. \citet{tao2023acoustic} introduced a self-attentional U-Net method to improve the noise resistance of the inversion process. \citet{ning2024transformer} proposed a hybrid approach combining the advantages of Transformer in capturing global features with CNN’s ability to extract local features, achieving high-resolution inversion results from seismic data.

The existing network architectures have reached a high level of maturity, and further improvement has been difficult to bring significant improvement. Therefore, the focus has shifted towards optimizing training methodologies and understanding their impact on model performance. In practical applications, the generalization ability of conventional supervised learning (SL) method is insufficient due to the limited number of logging labels \citep{yuan2022double, song2024reservoir}. To address this problem, \citet{alfarraj2019semi} proposed a semi-supervised learning (SSL) method based on the seismic convolution model. However, this method relies on the exact solution of seismic wavelets, so its applicability is poor \citep{wu2021semi}. \citet{shi2024seimic} proposed a semi-supervised learning workflow with closed-loop mapping, which effectively improved the accuracy and robustness of inversion results by mapping seismic data and impedance data in a closed-loop manner.

However, most learning methods rely on time domain seismic data as input, which limits the network to learn only fixed-bandwidth information, leading to insufficient resolution of the inversion results \citep{zhu2022data}. In addition, because the earth’s medium is complex and variable \citep{yuan2019impedance,wang2022propagating}, some thin layers and reservoirs may be hidden in the weak reflection areas of the seismic profile \citep{castagna2016phase}. It is difficult for the network to establish a mapping relationship between these weak signals and seismic impedance.

To solve these problems, we propose a time-frequency phase mixed domain closed-loop seismic inversion network (TFP-CSIN). Firstly, we construct a closed-loop seismic inversion network structure using bidirectional gated recurrent units (Bi-GRU) and CNN to establish the bidirectional mapping relationship between seismic data and acoustic impedance. Next, we develop a semi-supervised learning framework that utilizes a limited number of well-logging labels and near-well seismic data for supervised training, while incorporating a large number of non-near-well seismic data for unsupervised training. To enable the network to capture high-resolution information across the full frequency spectrum, we incorporate frequency domain information derived from the Fourier transform of seismic data. The instantaneous phase of seismic data is highly sensitive to complex structures and can enhance the detection of fluid content and subtle reservoir characteristics \citep{pei2022improved,de2019case}. To improve the detection of weak signals, we introduce phase domain information obtained through the Hilbert transform. Finally, we conduct numerical experiments using synthetic and field data to verify that the TFP-CSIN method outperforms traditional seismic inversion techniques in terms of resolution and thin layer identification.

\section{Theory and methodology}

\subsection{forward model}
According to the time domain convolution model proposed by \citep{robinson1967predictive}, Post-stack seismic data can be obtained by convolution of seismic wavelet and reflection coefficient, which is expressed as:
\begin{equation}\label{h1}
D(t) = W(t)R(t) + N(t),
\end{equation}
where $D(t)$ represents synthetic seismic data, $W(t)$ represents seismic wavelet, $R(t)$ represents seismic reflection coefficient sequence, $N(t)$ represents the noise in the measurements, and $t$ is two-way travel time.Similarly, the reflection coefficient can be generated by impedance, as follows:
\begin{equation}\label{h2}
 {R(t)} = \frac { {I} _ { t + 1 } - {I} _ { t } } { {I} _ { t + 1 } + {I} _ { t } },
\end{equation}
where $t + 1$ and $t$ represent the seismic impedance of layers $t+1$ and $t$, respectively. The corresponding synthetic seismic data can be calculated by combining equation~\ref{h1} and equation~\ref{h2}. 

\subsection{Time-frequency phase mixed domain inverse problem}

Equation~\ref{h1} is a forward equation in the time domain, which can also be applied to the frequency domain by the Fourier transform, and is expressed as:
\begin{equation}\label{h3}
F( \omega ) = \int _ { 0 } ^ { + \infty } D(\sigma) e ^ { - i \omega \tau(\sigma) } d\sigma + N(w), 
\end{equation}
where $F( \omega )$ represents the frequency spectrum of seismic data, $\tau(\sigma)$ shows the depth in the domain, $N(w)$ represents the spectrum of noise. Equation~\ref{h3} can be written in matrix form as:


\begin{equation}\label{h4}
\underbrace{
\begin{bmatrix} 
F( \omega_1 )\\ F( \omega_2 )\\\vdots \\F( \omega_m )
\end{bmatrix}}_{F}
=
\underbrace{\begin{bmatrix}   
e^{-i\omega_1\tau_1} & e^{-i\omega_1\tau_2} & \dots  & e^{-i\omega_1\tau_n}\\   
e^{-i\omega_2\tau_1} & e^{-i\omega_2\tau_2} & \dots  & e^{-i\omega_2\tau_n}\\ 
\vdots & \vdots & \ddots & \vdots \\   
e^{-i\omega_m\tau_1} & e^{-i\omega_m\tau_2} & \dots  & e^{-i\omega_m\tau_n}
\end{bmatrix}}_{G}
\underbrace{
\begin{bmatrix}D_1 \\D_2 \\\vdots \\D_n
\end{bmatrix}}_{D}
+
\underbrace{
\begin{bmatrix}
N(\omega_1)\\ N(\omega_2)\\ \vdots \\ N(\omega_n)
\end{bmatrix}}_{N_F}.
\end{equation}
Further, equation~\ref{h4} can be simplified as follows:
\begin{equation}\label{h5}
F = GD + N_F,
\end{equation}
where $F$ represents the frequency spectrum of seismic data, $G$ represents the forward operator in the frequency domain, $D$ represents synthetic seismic data, $N_F$ represents the spectrum of noise. Since equation~\ref{h5} is complex, The real and imaginary parts can be rewritten, respectively,
as:\begin{equation}\label{h6}
{F}' = {G}'D + N_F,
\end{equation}
where ${F}'=\begin{bmatrix}
real(F)  &
imag(F)
\end{bmatrix}^T$,
${G}'=\begin{bmatrix}
real(G)  &
imag(G)
\end{bmatrix}^T$,
$real(\ast)$ and $imag(\ast)$ represent real and imaginary parts, respectively. equation~\ref{h6} can be used to establish the frequency domain information of seismic data.

To improve the recognition ability and resolution of the weak reflection areas, the Hilbert transform (HT) is used to calculate the complex signal and instantaneous phase, which can be expressed as:
\begin{equation}\label{h7}
\tilde{D}(t)  = D(t) + i \hat{D}(t),
\end{equation}
where $\hat{D}(t)=Hilbert(D(t))$, and $Hilbert(D(t))$ represents the HT of the seismic signal. Thus, the instantaneous phase of the seismic data can be expressed as:
\begin{equation}\label{h8}
P = \tan^{-1}({D(t)}/{\hat{D}(t)}),
\end{equation}
where $P$ represents the instantaneous phase of the seismic data. Since the variable in
equation~\ref{h8} is the time domain reflectometry coefficient, it can be rewritten as:

\begin{equation}\label{h9}
P = KD + N_P,
\end{equation}
where $K$ denotes the phase domain forward operator, $N_P$ denotes the noise phase spectrum.


\subsection{Closed-loop seismic inversion network}
In this section, we further use the network to establish the objective function of time-frequency phase domain inverse problem. We provide a dataset comprising post-stack seismic data, which includes both labeled and unlabeled seismic data. These can be represented as $D_l=\left[D_{l}^1,D_{l}^2, \ldots ,D_{l}^{M-1}, D_{l}^{M}\right]$ and $D_u=\left[D_{u}^1,D_{u}^2, \ldots ,D_{u}^{N-1}, D_{u}^{N}\right]$,
where $D_{l}^{i}$ represents the labeled seismic data for $i$-th trace, $D_{u}^{i}$ represents the unlabeled seismic data for $i$-th trace, $M$ and $N$ are total number of labeled and unlabeled seismic traces, respectively. Simultaneously, we give well-logs data corresponding to $D_l$, which is expressed as $I=\left[I^1,I^2, \ldots ,I^{M-1}, I^{M}\right]$.

In supervised learning, we often use labeled seismic data and logging data to establish a nonlinear mapping relationship. The specific mathematical formula can be expressed as:
\begin{equation}\label{h10}
I^{\prime} \approx \mathcal{F}_{w}(D_l),
\end{equation}
where $I^{\prime}$ denotes the estimated acoustic impedance, $\mathcal{F}_{w}$ refers to the inversion network, represented as $\mathcal{F}_{w}= \left[ w_{\mathcal{F}}^{1},w_{\mathcal{F}}^{2}, \ldots ,w_{\mathcal{F}}^{L-1},w_{\mathcal{F}}^{L}\right]$, where $w_{\mathcal{F}}^{i}$ is the weight vector of the $i$-th layer in the network structure. By training the network and updating the weight vector $w$. However, the amount of logging data is often limited in practical geophysical applications, which constrains the generalization ability of supervised learning models. To maximize the utility of unlabeled seismic data, we implement a semi-supervised learning approach. In addition, it is difficult to obtain wavelet parameters using a forward convolutional model. Therefore, we use closed-loop neural network architecture as an alternative, which is expressed as:
\begin{equation}\label{h11}
{D_u}^{\prime} \approx \mathcal{H}_{w}(I),
\end{equation}
where ${D_u}^{\prime}$ denotes the estimated seismic data, $\mathcal{H}_{w}$ refers to the closed-loop network. Finally, we derive the objective function in the time-frequency mixed domain by integrating equations~\ref{h6}, ~\ref{h9}, ~\ref{h10}, and ~\ref{h11}. This function can be expressed as:
\begin{multline}\label{h12} 
J(\mathcal{F}_{w},\mathcal{H}_{w}) = 
\min_{\mathcal{F}_{w}} 
\underbrace{||{I_{l}}-\mathcal{F}_{w}(D_{l})||_2^2}_{Loss_{I}} +
\min_{\mathcal{H}_{w}} 
\underbrace{||{D_{l}}-\mathcal{H}_{w}(I_{l})||_2^2}_{Loss_{C}} +
\\
\min_{\mathcal{F}_{w},\mathcal{H}_{w}} 
\lambda_1\underbrace{||D_{U}-\mathcal{H}_{w}(\mathcal{F}_{w}(D_{U}))||_2^2}_{Loss_{T}}  + 
\min_{\mathcal{F}_{w},\mathcal{H}_{w}} 
\lambda_2\underbrace{||F_{U}-G^{\prime}\mathcal{H}_{w}(\mathcal{F}_{w}(D_{U}))||_2^2}_{Loss_{F}}  + 
\\
\min_{\mathcal{F}_{w},\mathcal{H}_{w}} 
\lambda_3\underbrace{||P_{U}-K\mathcal{H}_{w}(\mathcal{F}_{w}(D_{U}))||_2^2}_{Loss_{P}},
\end{multline}
where $\left \| \ast  \right \| _{2}^{2}$ denotes the L2 norm, $\lambda_1$, $\lambda_2$, and $\lambda_3$ denote the unsupervised learning weighting coefficients for the time domain, frequency domain, and phase domain, respectively. Finally, the loss function in training can be written as:
\begin{equation}\label{h13}
Loss= Loss_{I} + Loss_{C} + \lambda_1 Loss_{T} + \lambda_2 Loss_{F} + \lambda_3 Loss_{P},
\end{equation}
where $Loss$ is the total loss by the TFP-CSIN. $Loss_{I}$ represents the supervised loss of the inversion network, $Loss_{C}$ represents the supervised loss of the closed-loop network, $Loss_{T}$, $Loss_{F}$, and $Loss_{P}$ represent unsupervised losses in the time domain, frequency domain, and phase domain, respectively.

Figure~\ref{fig:figure1} shows the training process of the TFP-CSIN. First, labeled seismic data are fed into the seismic inversion network to generate predicted impedance data. simultaneously, well-log impedance is fed into the closed-loop network to generate reconstructed seismic data. The inversion network is trained by updating the supervised loss between the well-logs data and the predicted impedance. At the same time, the closed-loop network is also supervised trained by updating the losses between labeled seismic data and reconstructed seismic data. To improve the generalization of the network, a large amount of non-near-well seismic data are treated as unlabeled data and sequentially trained through the inversion network and closed-loop network during each iteration. Furthermore, to improve the inversion resolution and take into account the sensitivity of weak signals, the frequency information is obtained through the Fourier transform, while phase information is extracted via the Hilbert transform. These transformations are subsequently utilized to calculate unsupervised loss, which serves as a constraint in the inversion process.

In our experiments, we constructed a seismic inversion network and a closed-loop network with Bi-GRU and CNN architectures, respectively, as shown in Figure~\ref{fig:figure2}. Figure~\ref{fig:figure2}(a) shows the structure of the seismic inversion network. The network first inputs the seismic data into the three-layer Bi-GRU for sequence modeling. Additionally, considering the advantage of CNN in local feature extraction, we use three-layer CNN in parallel for feature extraction. Then, the sequence modeling information is spliced with the results of local feature extraction. Since the sampling rate of seismic data is lower than that of logging data in real situations, we add three-layer CNN to match the sampling rate. Finally, the network passes through a single CNN layer and a fully connected layer as a regression layer to match the seismic impedance sequence length. Figure~\ref{fig:figure2}(b) shows the architecture of the closed-loop regression network, consisting of two-layer CNN and one fully connected layer. This design is based on the following theoretical considerations: the synthesis process of wave impedance to seismic data can be regarded as a two-step convolution operation. The first is the convolution of acoustic impedance to reflection coefficients, and the second is the convolution of reflection coefficients to seismic data. The introduction of the fully connected layer ensures that the reconstructed seismic data maintains the same sequence length as the original seismic data. Both networks are synchronously optimized usingthe adaptive moment estimation (Adam) optimization algorithm \citep{diederik2014adam} and ReLU is used as the activation function \citep{nair2010rectified}. In each layer, we also add a group normalization (GN) structure to improve the training stability and convergence speed of the network \citep{wu2018group}.

\subsection{Transfer Learning}
Discrepancies between training data and field data can lead to degraded model performance when directly applied. Retraining the model with new real data is a potential solution. However, the limited amount of logging data makes it difficult to achieve a robust network through direct retraining. To address this challenge, we implement a transfer learning strategy. Figure~\ref{fig:figure3} shows the transfer learning process in TFP-CSIN. In our method, we maintain all layers of the pre-trained network derived from model data, except for the fully connected layer. Next, we fine-tune this modified pre-trained network using a training dataset constructed from log data.

First, we keep the layers of the pre-trained network from the model data fixed, except for the fully connected layer. We then fine-tune this pre-trained network using a training data set built from the log data. This method ensures the stability of the network and makes the updated network effectively applicable to real data.

\section{Experimental result} 
To verify the validity of the proposed method, we first determined the magnitude of the weighting coefficients for equation~\ref{h13}. Then, five different experimental groups were designed for ablation experiments to allow longitudinal comparison. All these networks use the same structure and architecture.
The following is a detailed description of the network training and testing process. To improve the stability of the training process and the speed of convergence, we normalized the input data. The normalization formula is expressed as: 
\begin{equation}\label{h14}
X^{\prime}= \frac{X- \mu_{X}}{\sigma_X},
\end{equation}
\begin{equation}\label{h15}
Y^{\prime}= \frac{Y- \mu_{Y}}{\sigma_Y},
\end{equation}
where $X^{\prime}$ and $Y^{\prime}$ denote the normalized seismic data and the normalized acoustic impedance data, respectively. $X$ and $Y$ represent the input seismic data and the acoustic impedance data, respectively. $\mu_{X}$ and $\mu_{Y}$ represent the mean values of $X$ and $Y$, while $\sigma_X$ and $\sigma_Y$ denote their respective standard deviations. 
To quantify the accuracy of the inversion results, we use three metrics commonly used in regression analysis. Namely, the Pearson correlation coefficient (PCC), the coefficient of determination ($\mathrm{R^{2}}$), and the signal-to-noise ratio (SNR). Both are defined as:
\begin{equation}\label{h16}
\mathrm{PCC}(y,\hat{y}) = \frac{\sum_{i=1}^{M}\sum_{j=1}^{N}(y_{i,j}- \mu _{y})(\hat{y}_{i,j} - \mu _{\hat{y}})}{\sqrt{\sum_{i=1}^{M}\sum _{j=1}^{N}(y_{i,j}- \mu _{y})^{2}}\sqrt{\sum_{i=1}^{M}\sum _{j=1}^{N}(\hat{y}_{i,j}- \mu _{\hat{y}})^{2}}},
\end{equation}
\begin{equation}\label{h17}
\mathrm{R^{2}}(y,\hat{y}) = 1- \frac{\sum_{i=1}^{M}\sum _{j=1}^{N}(y_{i,j}- \hat{y}_{i,j})^{2}}{\sum_{i=1}^{M}\sum _{j=1}^{N}(y_{i,j}- \mu _{y})^{2}},
\end{equation}
\begin{equation}\label{h18}
\mathrm{SNR}(y,\hat{y}) = 10\log_{10}{\frac{\sum_{i=1}^{M}\sum _{j=1}^{N}y_{i,j}^{2}}{\sum_{i=1}^{M}\sum _{j=1}^{N}(y_{i,j}- \hat{y}_{i,j})^{2}}} ,
\end{equation}
where $y$ and $\hat{y}$ are the real model and the predicted results, respectively. $\mu_{y}$ and $\mu_{\hat{y}}$ denote the average of the $y$ and $\hat{y}$, respectively. $M$ is the number of rows in the matrix and $N$ is the number of columns in the matrix.

\subsection{Tests on marmousi2 model} 
To train our model, we first selected a part of data from the Marmousi2 post-stack dataset \citep{martin2006marmousi2}. This dataset was simulated using a zero-phase Ricker wavelet with a dominant frequency of 20 Hz and obtained through forward modeling based on the equation~\ref{h1}. As shown in Figure~\ref{fig:figure4}(a), the model contains 3000 traces and 690 time samples with a time interval of 2 ms. The seismic traces were then downsampled by a factor of 6 to simulate the difference in resolution between the seismic and well-logs data. In the Marmousi2 model of 3000 traces, there are 18 traces with known acoustic impedance as labeled data, and the number was much less than $1\%$ of the total number, as shown in Figure~\ref{fig:figure4}(b). In the network process, the hyperparameters for TFP-CSIN are set as follows: the number of training epochs is 500, the initial learning rate to 0.001 and the weight decay to 0.0001, dropout to 0.2, and the batch size is 40. The experiments were conducted on a computer running the Windows 11 operating system. The system was equipped with 32 GB of RAM, an Intel Core i5-12490F processor, and an NVIDIA GeForce GTX 3060 graphics processing unit (GPU). 

Figure~\ref{fig:figure5} shows the curve of each type of loss defined in equation~\ref{h13} over 500 training iterations. The vertical axis shows the total loss on the training data set. Note that the losses converge steadily, albeit at different scales. Therefore, determining the weighting coefficients for each type of loss is critical. To determine the magnitudes of the weight coefficients $\lambda_1$, $\lambda_2$, and $\lambda_3$ in equation~\ref{h13}, we designed three sets of ablation experiments: 

1) time domain SSL (T-SSL) method ($\lambda_2=\lambda_3=0$).

2) frequency domain SSL (F-SSL) method ($\lambda_1=\lambda_3=0$).

3) phase domain SSL (P-SSL) method ($\lambda_1=\lambda_2=0$).

Figure~\ref{fig:figure6} shows the change curves of the three types of typical indicators ($\mathrm{PCC}$, $\mathrm{R^{2}}$, and $\mathrm{SNR}$) for each method. It can be seen that the evaluation coefficients first increase and then decrease with the gradual increase of the weight coefficient $\lambda$. The rectangular line box indicates the maximum value of the evaluation coefficient. The results indicate that the optimal weight for the T-SSL method is $\lambda_1=1$, the optimal weight for the F-SSL method is $\lambda_2=1\times10^{-4}$, and the optimal weight for the P-SSL method is $\lambda_3=1\times 10^{-1}$. 

We compare the performance of the proposed method (TFP-CSIN) with four existing approaches: SL, T-SSL, F-SSL, and P-SSL. Figure~\ref{fig:figure7}(a)-(e) shows the impedance profiles obtained using these different methods. The SL method shows low resolution and poor lateral continuity, as shown in Figure~\ref{fig:figure7}(a). In contrast, the T-SSL, F-SSL, and P-SSL methods significantly improve the resolution of the inversion, although they show certain deficiencies in detail preservation. The TFP-CSIN method outperforms the others in terms of both lateral continuity and resolution. Figure~\ref{fig:figure7}(f)-(j) shows the residuals of inversion results for each method, with the TFP-CSIN method showing the smallest residuals. In regions of zero amplitude, the TFP-CSIN method effectively preserves the anti-interference properties of the T-SSL method, ensuring longitudinal continuity in the inversion results (highlighted by the red frame in Figure~\ref{fig:figure7}(g) and (j)). In areas of strong amplitude variation, the TFP-CSIN method inherits the high-resolution capabilities of the F-SSL method (indicated by the blue frame in Figure~\ref{fig:figure7}(h) and (j)). Furthermore, the TFP-CSIN method retains the weak signal sensitivity of the P-SSL method, thereby enhancing the detection of weak reflection areas, as shown in the purple frame in Figure~\ref{fig:figure7}(i) and (j). The quantitative results of all traces not used in the training are summarized in Table~\ref{tbl:table1}. The quantitative results show that the TFP-CSIN inversion results match the true model better than other methods. Figure~\ref{fig:figure8} shows the inversion results of traces 430 and 2300, which are not included in the labeled data. It can be seen that P-SSL and TFP-CSIN have the highest prediction accuracy of the inversion results, regardless of whether trace 246 or trace 560 is chosen for comparison. The enlarged plot on the right shows a slight superiority of TFP-CSIN over P-SSL. The SL, T-SSL, and P-SSL methods exhibit larger variations in absolute impedance values, resulting in large inversion biases, as shown by the regions marked with arrows in Figure~\ref{fig:figure8}. This phenomenon is due to the variation in absolute impedance values and the constraints imposed by frequency bandwidth limitations. 

Figure~\ref{fig:figure9} shows the normalized amplitude spectra of each method at trace 430 and trace 2300. The box in Figure~\ref{fig:figure9}(a) highlights the low-frequency components in the 0-10 Hz range. The SL method demonstrates limited effectiveness in recovering these low-frequency components, while the other methods exhibit improved performance. The T-SSL and F-SSL methods show varying effectiveness across different frequency bands. It can be seen that they can accurately detect high frequency signals, but struggle with the high frequency components in Figure~\ref{fig:figure9}(b). Notably, the F-SSL and T-SSL methods effectively capture a wider frequency range at both traces 430 and 2300, including both low and high frequencies.

To verify the noise resistance and feasibility of the TFP-CSIN method, we add different levels of Gaussian noise to the original data using equation~\ref{h18} and obtain signals with SNRs of 5, 10, and 20 dB. Figure~\ref{fig:figure10}(a)-(c) shows the seismic data with different SNRs. To visualize the noise level, a single random trace was extracted and shown in Figure~\ref{fig:figure10}(d) to show the waveform with different SNRs. The impedance inversion results for different noise backgrounds are shown in Figure~\ref{fig:figure11}(a)-(c), Figure~\ref{fig:figure11}(d)-(f), Figure~\ref{fig:figure11}(g)-(i), Figure~\ref{fig:figure11}(j)-(l), and Figure~\ref{fig:figure11}(m)-(o) for the SL, T-SSL and F-SSL, P-SSL, and TFP-CSIN, respectively. The inversion results of TFP-CSIN are more stable and have better continuity than other methods. Moreover, the inversion results of TFP-CSIN reveal more details in some localized regions (marked with black arrows). To visually and quantitatively compare the degree of noise impact of different methods, $\mathrm{PCC}$, $\mathrm{R^{2}}$, and $\mathrm{SNR}$ were used to calculate the fitting value, as shown in Figure~\ref{fig:figure12}. With the increase of noise, we can get the following information from Figure~\ref{fig:figure11} (marked with black rectangles) and Figure~\ref{fig:figure12}. It can be seen that the T-SSL inversion results can remain maintain good stability. For the frequency domain inversion results, even when the resolution is high, the inversion results are unstable with the increase of noise. Although phase domain inversion effectively highlights weak reflection areas, its performance is limited in detecting strong amplitude features.

\section{Field Data Example} 
Based on the success of the TFP-CSIN method on synthetic model data, we applied it to real data from the Yingqiong basin in South China Sea. The lithology of study area is mainly composed of fine sandstone and medium sandstone. There are many thin sand bodies in the working area. Figure~\ref{fig:figure13} shows a horizontal map of the original post-stack seismic data consisting of 551 inlines and 901 crossline. The purple dotted line indicates (inline=356) the location of the vertical well profile used to test the inversion results. The information from well w4 (inline=356 and crossline=845) is used for the blind test, and the remaining 8 wells were used for fine-tuning the network. 

Figure~\ref{fig:figure14}(a) shows the profiles of the original post-stack seismic data (obtained from the purple dashed lines shown in Figure~\ref{fig:figure13}). The depth sampling range for the displayed data is 1350 to 1750 ms, with a sampling interval of 2 ms. In addition, this profile includes the training of 2 wells. Figure~\ref{fig:figure14}(b) and Figure~\ref{fig:figure14}(c) show the wave impedance curves of wells W1 (crossline=533) and W8 (crossline=1032), respectively. Due to the fact that the high-frequency logging to calibrate low-frequency seismic data, the predicted result will differ from the actual result. Therefore, we use a low-pass filter to smooth the well curve. The well logging data after filtering is shown in Figure~\ref{fig:figure14}.

Due to lithologic characteristics of the study area, there are many thin sand layers and weak reflection areas, so the inversion method with higher resolution is needed to accurately identify them. Figure~\ref{fig:figure15}(a) shows the inversion results of SL without transfer learning, which lack continuity and show obvious blurrings. Figure~\ref{fig:figure15}(b) shows the inversion results of SL with transfer learning conditions, and the lateral continuity of inversion results is significantly improved. Similarly, Transfer learning strategies are also applied in semi-supervised methods. Figure~\ref{fig:figure15}(c)-(f) shows the inversion results of different semi-supervised methods. Compared with the SL method, all four semi-supervised methods achieve higher resolution. However, there are differences in the details. In terms of thin sand layer recognition, T-SSL, F-SSL, and TFP-CSIN achieve satisfactory performance (marked with black arrows). TFP-CSIN shows superior performance in horizon identification (marked with dashed ellipses). The P-SSL and TFP-CSIN methods demonstrate optimal performance in delineating weak reflection areas (marked with red arrows). 

Furthermore, We conducted well bypass curve analysis for well W4, and the results are shown in Figure~\ref{fig:figure16}. In comparison to other inversion method, the impedance inversion result of TFP-CSIN are closer to the actual logging curve. Table~\ref{tbl:table2} shows the $\mathrm{PCC}$, $\mathrm{R^{2}}$ and $\mathrm{SNR}$ between the inversion results at the location of the well W4 and the well-log curves. It can be seen that the $\mathrm{PCC}$, $\mathrm{R^{2}}$ and $\mathrm{SSIM}$ indexes of the TFP-CSIN inversion results are the best. Therefore, compared with the semi-supervised inversion method in a single domain, the TFP-CSIN method in the time-frequency phase mixed domain can significantly improve the resolution of the inversion result. 


\section{Discussion}
Various deep neural networks have been introduced to improve the resolution of seismic inversion. However, the absence of physical constraints in these networks limits their applicability to complex geological formations, such as thin layers and hidden reservoirs. Enhancing the network structure alone is insufficient to significantly improve inversion resolution. In this paper, the time-frequency phase mixed domain inversion method gives consideration to the advantages of both time domain, frequency domain, and phase domain inversion. This method not only ensures robust noise resistance but also improves the resolution and detection capabilities for thin layers and areas of weak reflections. 

In this paper, the inversion objective function (equation~\ref{h13}) contains weighting coefficients $\lambda_1$, $\lambda_2$ and $\lambda_3$, and their influence on the inversion results is briefly described below. Coefficients $\lambda_1$, $\lambda_2$, and $\lambda_3$ are associated with the SNR, the resolution of the inversion results, and the weak reflection areas, respectively. The coefficient $\lambda_1$ should be increased when the actual data shows high noise levels. Conversely, $\lambda_2$ should be increased when the noise level is low. The coefficient $\lambda_3$ should be increased when there are numerous weak reflection areas in the field data.

\section{Conclusion}

In this paper, we propose a semi-supervised inversion method called TFP-CSIN to improve the resolution of inversion results and the ability to identify weak reflection areas. The Fourier transform captures frequency information within the closed-loop seismic inversion network, enabling the neural network to process data across the frequency bands. In addition, the Hilbert transform is used to impose phase constraints, which increases the sensitivity of the network to weak reflection regions. Synthetic data experiments show that the proposed method effectively improves the inversion resolution and the capability to identify weak reflection areas. Experiments on the field data further validate the effectiveness and practicability of the proposed method.

\section{Data and materials availability}
The original contributions presented in the study are included in the paper; further inquiries can be directed to the corresponding author.

\bibliographystyle{seg}  
\bibliography{TFP-CSIN}

\tabl{table1}{Quantitative evaluation of different methods for the synthetic data.}
{
\renewcommand\arraystretch{0.7}
\centering
\begin{center}
\begin{tabular}{c|c|c|c}
\hline \hline
\textbf{Model} & \textbf{PCC}    & \textbf{R\textasciicircum{}2} & \textbf{SNR}     \\ \hline
SL             & 0.9687          & 0.9241                        & 21.4393          \\ \hline
T-SSL          & 0.9779          & 0.9419                        & 22.6739          \\ \hline
F-SSL          & 0.9783          & 0.9482                        & 22.8394          \\ \hline
P-SSL          & 0.9797          & 0.9430                        & 22.9574          \\ \hline
TFP-CSIN        & \textbf{0.9828} & \textbf{0.9584}               & \textbf{24.0275} \\ \hline \hline
\end{tabular}
\end{center}
}

\tabl{table2}{Performance of different inversion results at the location of the well W4 in the field data.}
{
\renewcommand\arraystretch{0.7}
\centering
\begin{center}
\begin{tabular}{c|c|c|c}
\hline \hline
\textbf{Model} & \textbf{PCC}    & \textbf{R\textasciicircum{}2} & \textbf{SNR}     \\ \hline
SL             & 0.8564          & 0.7812                        & 18.8375          \\ \hline
T-SSL          & 0.9121          & 0.8293                        & 19.8814          \\ \hline
F-SSL          & 0.8826         & 0.8486                       & 21.0442          \\ \hline
P-SSL          & 0.8768          & 0.8388                        & 19.9785          \\ \hline
TFP-CSIN        & \textbf{0.9242} & \textbf{0.8613}               & \textbf{21.6280} \\ \hline \hline
\end{tabular}
\end{center}
}

\renewcommand{\figdir}{} 

\plot{figure1}{width=\textwidth}{The training process of the TFP-CSIN.}

\newpage
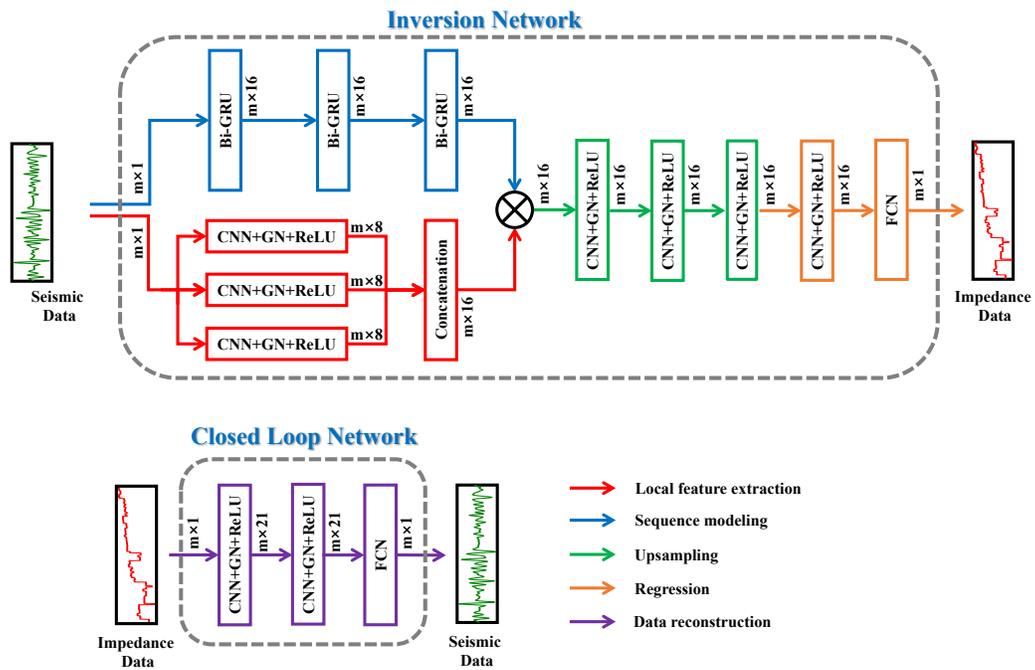
\plot{figure2}{width=\textwidth}{The structure of the inversion network and closed-loop network.}

\newpage
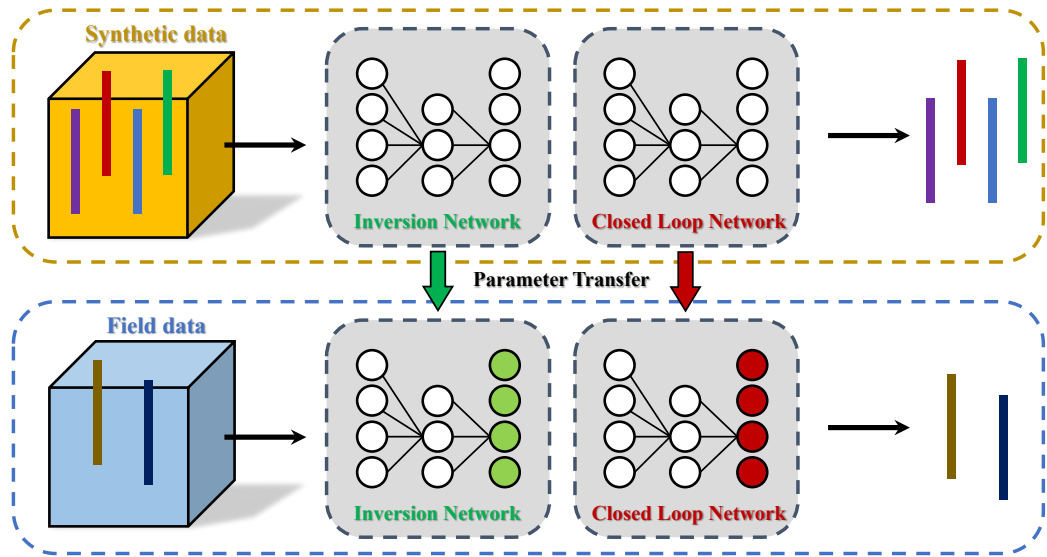
\plot{figure3}{width=\textwidth}{Schematic of transfer learning in TFP-CSIN. The white circles represent the network parameters derived from the synthesized data. The green circles denote the inversion network parameters fine-tuned during training with field data. The red circles indicate the closed-loop network parameters similarly fine-tuned using field data.}

\newpage
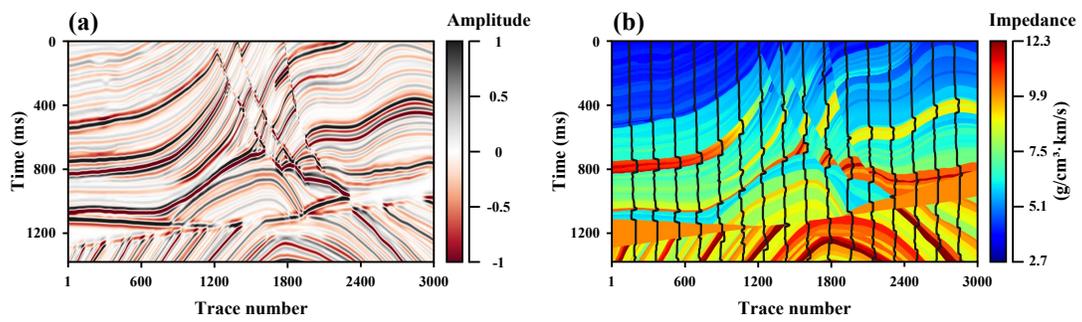
\plot{figure4}{width=\textwidth}{The seismic data and impedance data of the marmousi2 model. (a) Synthetic seismic profile. (b) True impedance profile. The solid black lines indicate the 18 traces that were extracted at equal intervals.}

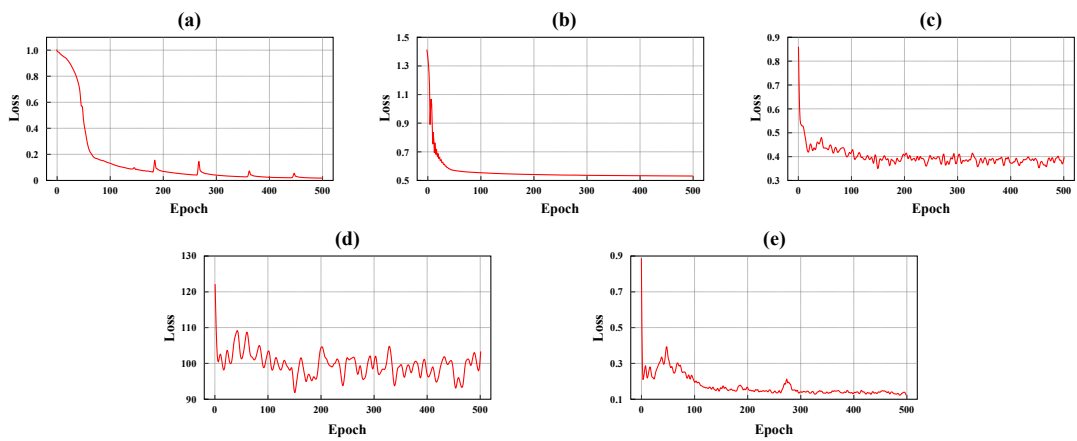
\plot{figure5}{width=\textwidth}{Loss function curves of (a) $Loss_I$, (b) $Loss_C$, (c) $Loss_T$, (d) $Loss_F$, and (e) $Loss_P$.}

\newpage
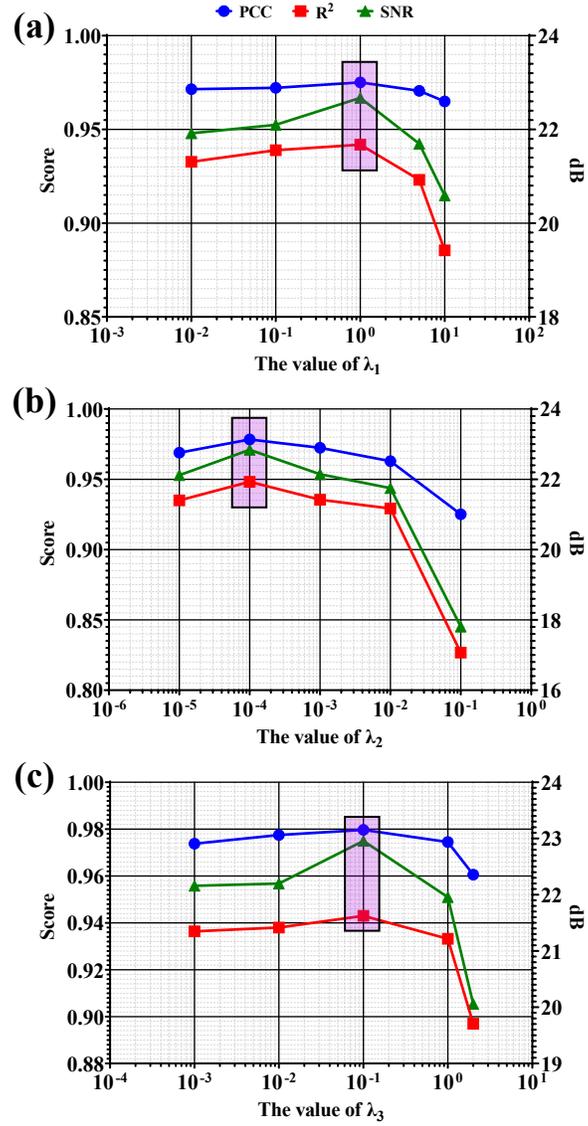
\plot{figure6}{width=0.6\textwidth}{Variation of PCC, $\mathrm{R^{2}}$, and SNR curves for each method with different weighting factors. (a) time domain SSL method, (b) frequency domain SSL method, and (c) phase domain SSL method. The rectangular boxes in the figure indicate the maximum values of the curves.}

\newpage
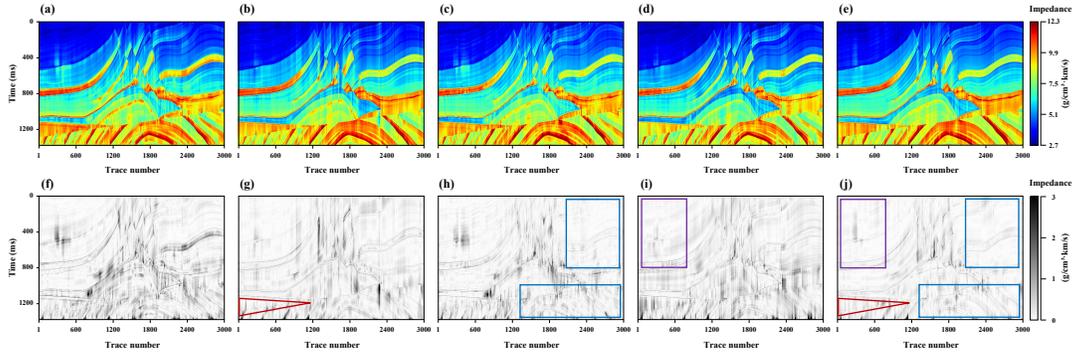
\plot{figure7}{width=1\textwidth}{Impedance inversion results of (a) the SL method, (b) the T-SSL method ($\lambda_1=1$, $\lambda_2=0$, $\lambda_3=0$), (c) the F-SSL method ($\lambda_1=0$, $\lambda_2=1\times10^{-4}$, $\lambda_3=0$), (d) the P-SSL method ($\lambda_1=0$, $\lambda_2=0$, $\lambda_3=1\times10^{-1}$), and (e) the TFP-CSIN method ($\lambda_1=5\times10^{-1}$, $\lambda_2=1\times10^{-5}$, $\lambda_3=2\times10^{-2}$). The lower panels (f)-(j) display the absolute error between the true model and the inversion results for each corresponding method.}

\newpage
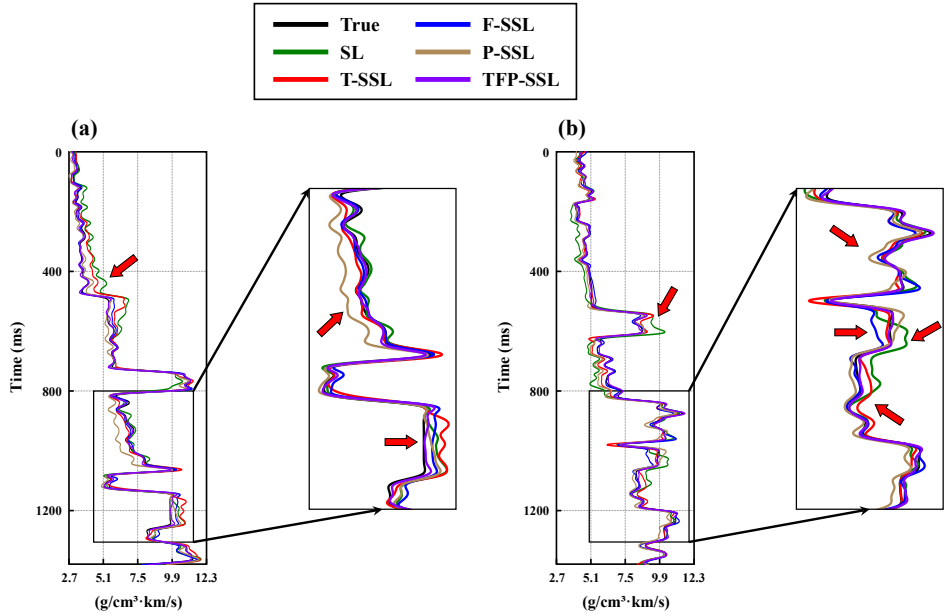
\plot{figure8}{width=\textwidth}{Inversion results at (a) trace 430 and (b) trace 2300 on the synthetic data. The black lines indicate true data, the green lines indicate the SL inversion results, the red lines indicate the T-SSL inversion results, the blue lines indicate the F-SSL inversion results, the brown lines indicate the P-SSL inversion results, and the purple lines indicate the TFP-CSIN inversion results.}

\newpage
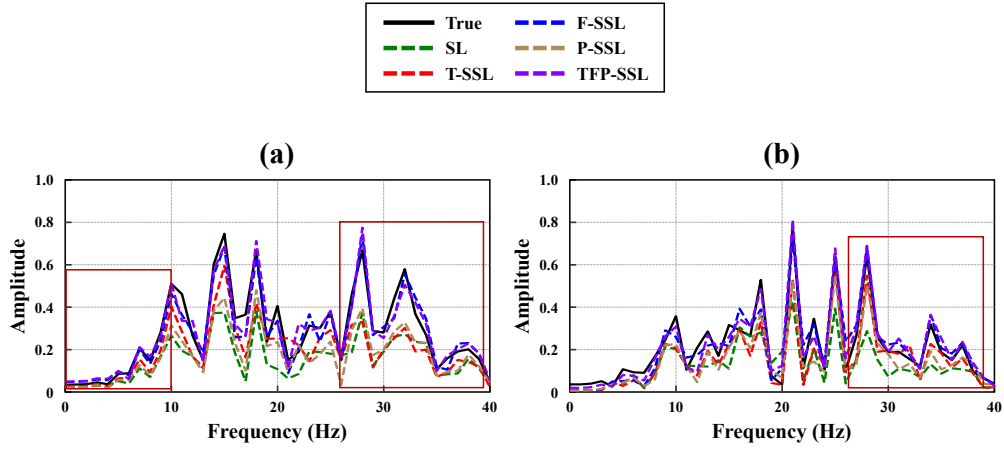
\plot{figure9}{width=\textwidth}{The normalized amplitude spectra of each method at (a) trace 430 and (b) trace 2300. The black lines indicate true data, the green dotted lines indicate the SL inversion results, the red dotted lines indicate the T-SSL inversion results, the blue dotted lines indicate the F-SSL inversion results, the brown dotted lines indicate the P-SSL inversion results, and the purple dotted lines indicate the TFP-CSIN inversion results.}

\newpage
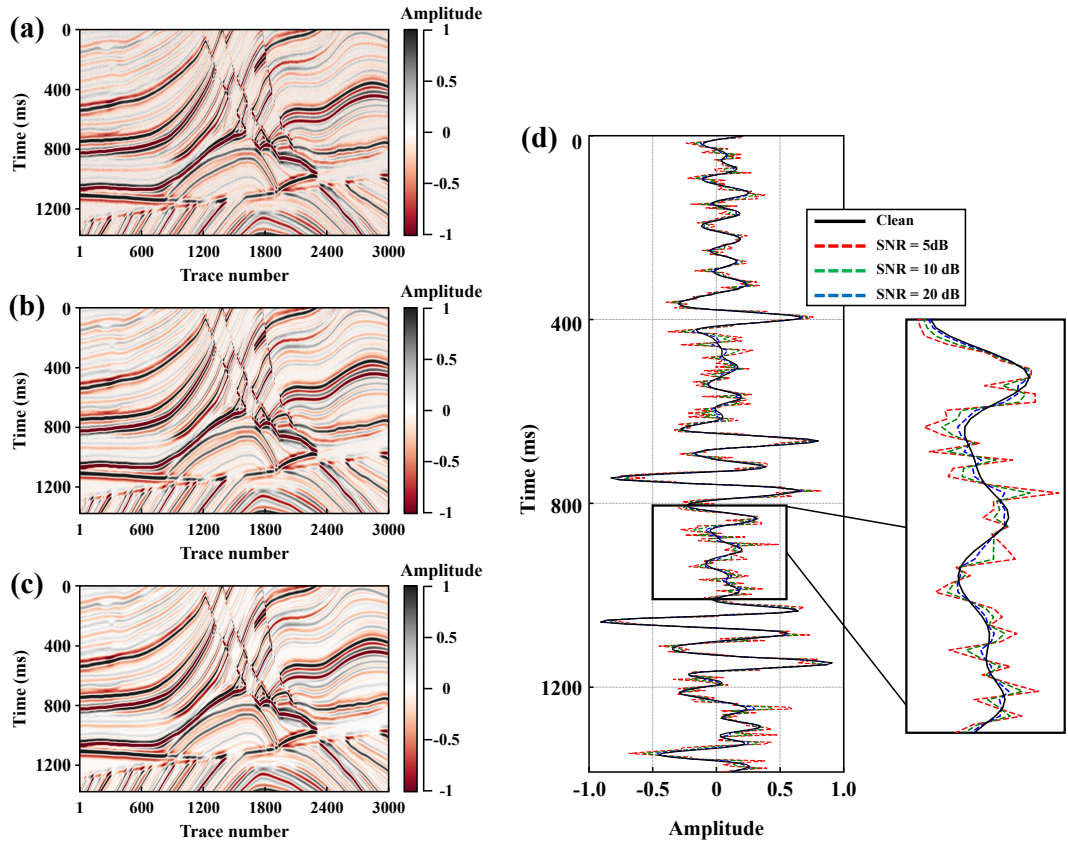
\plot{figure10}{width=\textwidth}{Seismic profiles with different SNRs. (a) SNRs = 5db, (b) SNRs = 10 dB, and (c) SNRs = 20 dB. (d) Waveform diagram of a randomly extracted trace.}

\newpage
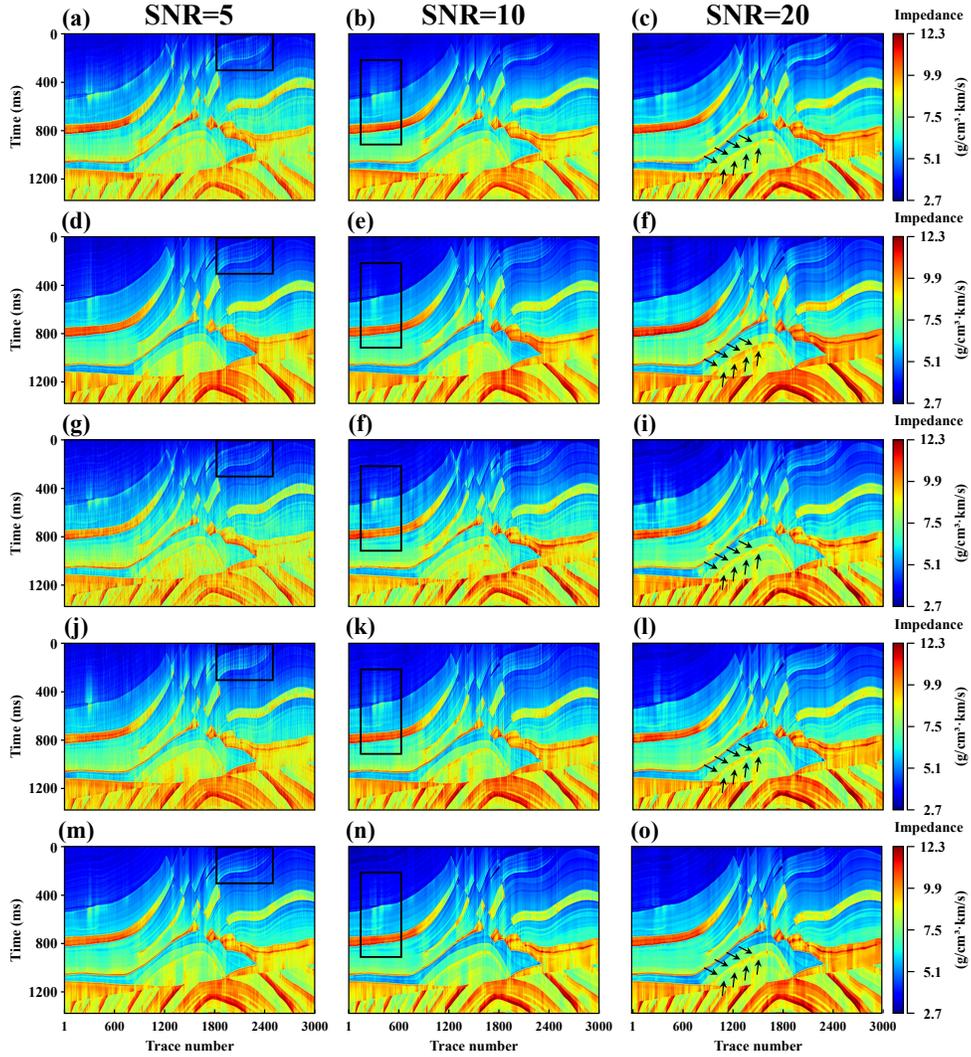
\plot{figure11}{width=0.9\textwidth}{Impedance inversion results of different methods under different noise conditions. (a)–(c) The SL method, (d)–(f) the T-SSL method ($\lambda_1=1$, $\lambda_2=0$, $\lambda_3=0$), (g)–(i) the F-SSL method ($\lambda_1=0$, $\lambda_2=1\times10^{-4}$, $\lambda_3=0$), and (j)–(i) the P-SSL method ($\lambda_1=0$, $\lambda_2=0$, $\lambda_3=2\times10^{-1}$). TFP-CSIN method results for: (m) SNR = 5dB ($\lambda_1=8\times10^{-1}$, $\lambda_2=8\times10^{-7}$, $\lambda_3=5\times10^{-3}$); (n) SNR = 10dB ($\lambda_1=7\times10^{-1}$, $\lambda_2=2\times10^{-6}$, $\lambda_3=7\times10^{-3}$); and (o) SNR = 20dB ($\lambda_1=6\times10^{-1}$, $\lambda_2=5\times10^{-6}$, $\lambda_3=1\times10^{-2}$).}
\newpage
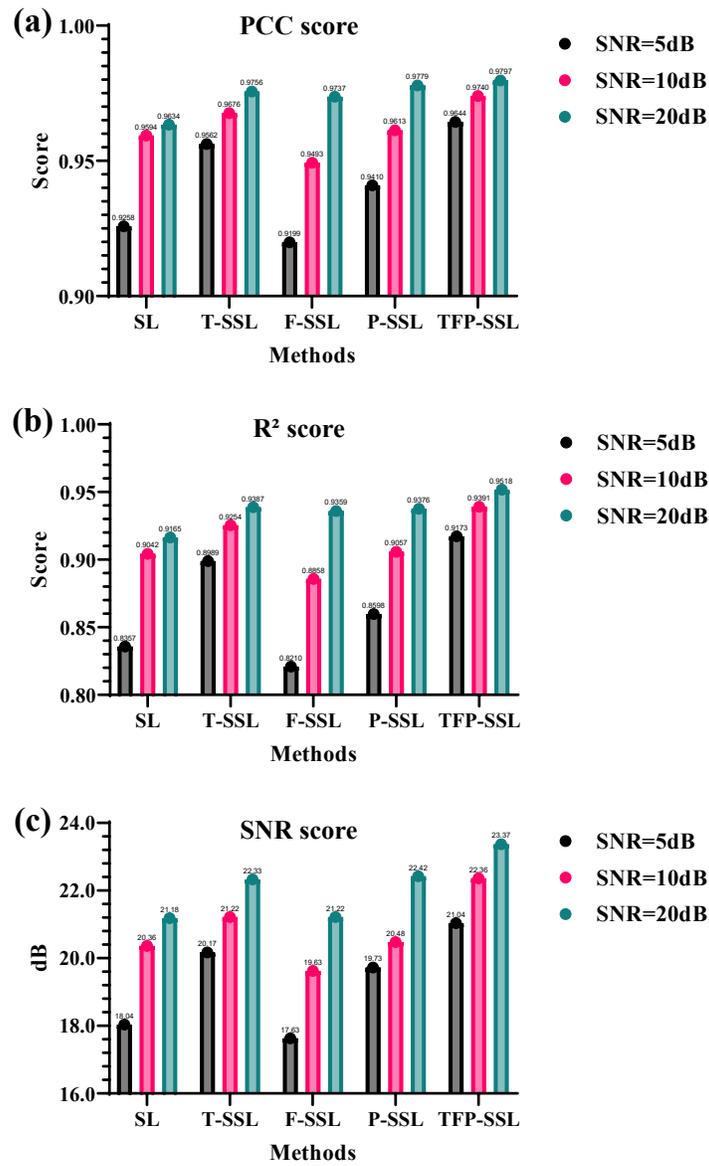
\plot{figure12}{width=0.7\textwidth}{Histograms of evaluation coefficients of inversion results from different methods applied to synthetic seismic data at different SNRs. (a) PCC, (b) $\mathrm{R^{2}}$, and (c) SNR.}

\newpage
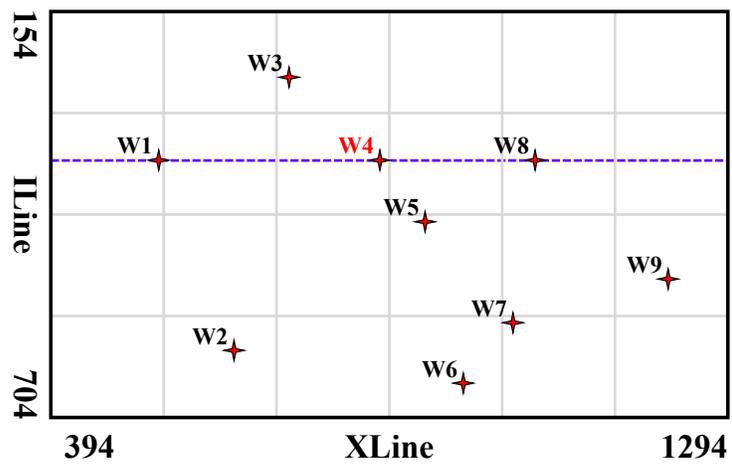
\plot{figure13}{width=\textwidth}{Spatial distribution of field data and the well logs. The purple dashed line (inline=356) indicates the location of the vertical well profile used to show the inversion results.}

\newpage
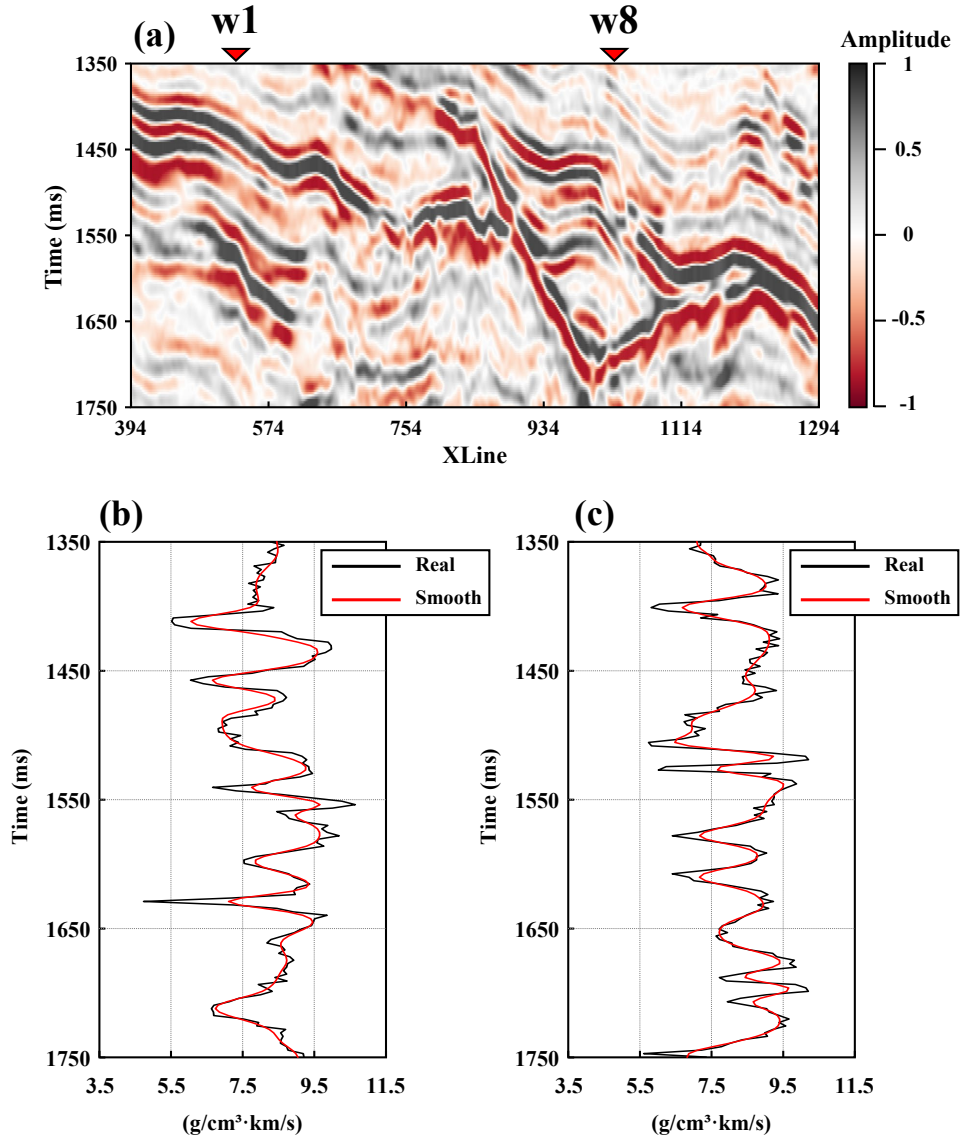
\plot{figure14}{width=\textwidth}{(a) Post-stack seismic data of the field. (b) Low-pass filtering of impedance curves for well W1. (c) Low-pass filtering of impedance curves for well W8.}

\newpage
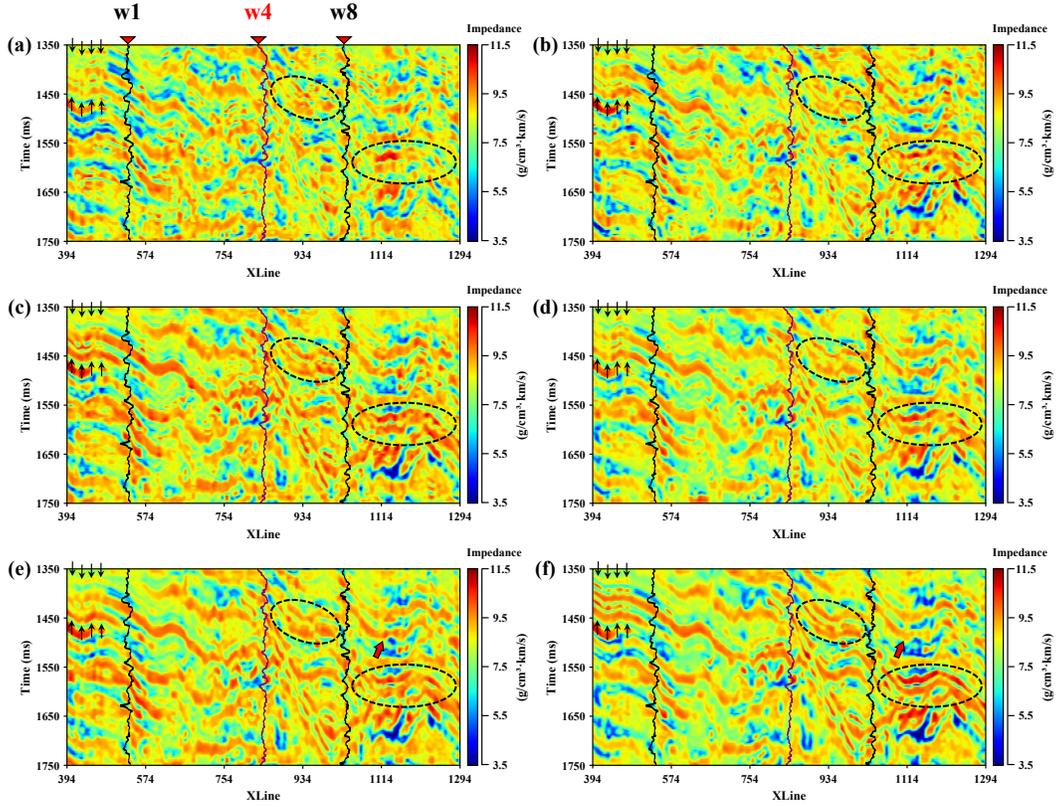
\plot{figure15}{width=\textwidth}{Comparison of inversion results from the different methods for the seismic profile (InLine=356). (a) the SL method without transfer learning, (b) the SL method with transfer learning, (c) the T-SSL method with transfer learning ($\lambda_1=1$, $\lambda_2=0$, $\lambda_3=0$), (d) the F-SSL method with transfer learning ($\lambda_1=0$, $\lambda_2=1\times10^{-4}$, $\lambda_3=0$), (e) the P-SSL method with transfer learning ($\lambda_1=0$, $\lambda_2=0$, $\lambda_3=2\times10^{-1}$), and (f) the TFP-CSIN method with transfer learning ($\lambda_1=7\times10^{-1}$, $\lambda_2=2\times10^{-5}$, $\lambda_3=3\times10^{-2}$).}

\newpage
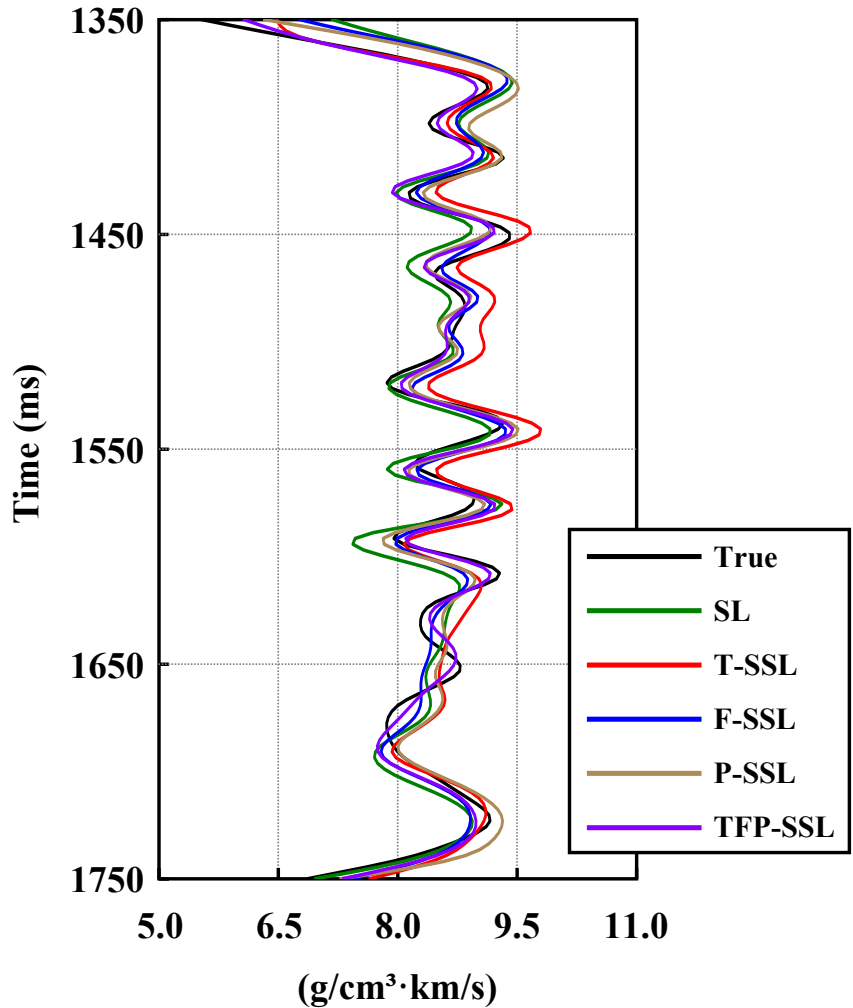
\plot{figure16}{width=\textwidth}{Impedance inversion results at the location of the well W4 in the field data. The black lines indicate the true data, the green lines indicate the SL inversion results, the red lines indicate the T-SSL inversion results, the blue lines indicate the F-SSL inversion results, the brown lines indicate the P-SSL inversion results, and the purple lines indicate the TFP-CSIN inversion results.}
\end{document}